\documentclass[%
 reprint,
nofootinbib,
 amsmath,amssymb,
 aps,
]{revtex4-2}

\usepackage[normalem]{ulem}
\usepackage{comment}
\usepackage[super]{nth}
\usepackage{graphicx}
\usepackage{dcolumn}
\usepackage{bm}
\usepackage[dvipsnames]{xcolor}
\usepackage{pgfplotstable, booktabs}
\usepackage{array}

\usepackage{hyperref}

\begin{document}

\preprint{APS/123-QED}

\title{High-energy interactions of charged black holes in full general relativity I: Zoom-whirl orbits and universality with the irreducible mass}

\author{M A. M. Smith} 
\email{mamsmith@arizona.edu, they/them/theirs}
\affiliation{%
 Department of Physics, The University of Arizona, Tucson, AZ, USA
}%
\author{Vasileios Paschalidis}
\email{vpaschal@arizona.edu}
\affiliation{Department of Astronomy, The University of Arizona, Tucson, AZ, USA
}
\affiliation{Department of Physics, The University of Arizona, Tucson, AZ, USA}

\author{Gabriele Bozzola}
\affiliation{Division of Geological and Planetary Science, California Institute of Technology, Pasadena, CA, USA}
\affiliation{Department of Astronomy, The University of Arizona, Tucson, AZ, USA}

\date{\today}

\begin{abstract}
We simulate high-energy scattering of equal-mass, nonspinning black holes
endowed with like charges in full general relativity while varying the impact parameter $b$. We show that electrodynamics does not suppress zoom-whirl orbits for at least charge-to-mass ratios $\lambda = 0.1, 0.4, 0.6$. However, we find that as $\lambda$ increases, the immediate merger and scattering thresholds defining the zoom-whirl regime move to smaller impact parameter $b/M_{\rm ADM}$, with $M_{\rm ADM}$ designating the binary black hole gravitational mass. This demonstrates that charge leaves observable imprints in key properties at energy scales where charge has negligible influence in head-on collisions. Additionally, we find that these threshold impact parameters become universal, i.e., charge-independent, when we normalize $b$ by the sum of the initial BH irreducible masses in the binary ($b/M_{\rm irr}$). This is the first explicit demonstration that the irreducible mass, which is proportional to the black hole areal radius, defines a fundamental gauge-invariant length scale governing horizon scale scattering events in the strong-field, dynamical spacetime regime.
\end{abstract}

\maketitle


\section{Introduction}

Numerical relativity simulations of high-energy black hole (BH) collisions have been used to investigate cosmic censorship, the maximum luminosity of physical processes, and the effect of internal BH structure, trans-Plankian scattering and AdS/CFT to name a few (see, e.g.,~\cite{cardoso_exploring_2015} for a review). Past simulations have explored the gamut of parameter space from head-on collisions of equal-mass, uncharged, non-spinning BHs \cite{sperhake_high-energy_2008, healy_high_2016} to general impact parameter \cite{shibata_high-velocity_2008, sperhake_cross_2009}, mass ratio \cite{sperhake_gravity-dominated_2016}, spin \cite{sperhake_superkicks_2011, sperhake_universality_2013}, and higher spacetime dimensions \cite{cook_black-hole_2017,sperhake_high-energy_2019,andrade_evidence_2022} (see \cite{cardoso_exploring_2015} for a comprehensive review and motivation of such fundamental physics studies). Recently, a new facet of these investigations has reopened: BH charge.

While numerical relativity simulations of charged BH binaries have been performed before, previous initial data have either been constraint-violating or confined to a moment of time-symmetry \cite{alcubierre_einstein-maxwell_2009,liebling_electromagnetic_2016}, restricting studies to head-on collisions from rest~\cite{zilhao_collisions_2012, zilhao_collisions_2014} (for a summary of such studies see \cite{bozzola_initial_2019}). The recent completion of a new Bowen-York-type initial data solver for binary BHs (BBHs) with arbitrary linear and angular momenta \cite{bozzola_initial_2019} now allows us to perform generic high-energy collisions of black holes endowed with U(1) charge (see \cite{mukherjee_conformally_2022} for a different approach to initial data for charged black holes). Since then, the Bowen-York-type initial data solver has been used to investigate high-energy, head-on collisions of charged BHs \cite{bozzola_does_2022}, and charged quasicircular binaries \cite{bozzola_general_2021, bozzola_numerical-relativity_2021, luna_kicks_2022, bozzola_can_2023}. Our work is intended to continue to fill out this uncharted region of the BBH parameter space by performing simulations of high-energy, charged BBHs near the scattering regime.

The scattering regime is defined by a critical impact parameter \(b_{\rm scat}\) that separates binaries that ultimately merge and those that scatter. In addition, an immediate merger threshold $b^*$separates the impact parameters that result in binaries merging within the first encounter ($b < b^*$) from impact parameters that result in multiple BH encounters ($b^* < b < b_{\rm scat}$) before merging~\cite{pretorius_black_2007, sperhake_cross_2009}.
Previous studies without charge have found that binaries at or near this scattering threshold provide new extremes. 

A uniquely relativistic phenomenon arises in the dynamics for $b^* < b < b_{\rm scat}$: zoom-whirl orbits, e.g.~\cite{bombelli_chaos_1992,levin_gravity_2000,glampedakis_zoom_2002,pretorius_black_2007,healy_zoom-whirl_2009,sperhake_cross_2009,gold_radiation_2010,gundlach_critical_2012}.  The zoom-whirl orbit is categorized by close, nearly-circular ``whirls" followed by wide, elliptical ``zooms." When exhibiting a zoom-whirl orbit, the BHs of a binary experience more than a single close encounter prior to merger. These orbits are a manifestation of the strong-field effects predicted by Einstein's theory of general relativity, and studying them can help improve our understanding of dynamical spacetime. Probing the high-energy scattering threshold regime with charged BHs pushes general relativity and electromagnetism to their limits, enabling us to understand Einstein-Maxwell theory in a regime that is generally not accessible with traditional theoretical methods (but see~\cite{PhysRevD.107.064051}). There are a number of unanswered questions regarding high-energy scattering of charged BBHs, and these questions cannot be answered by pure thought or earlier studies, since charge has not previously been considered.

In this work, we tackle questions about zoom-whirl orbits in charged BBHs. How does charge affect the dynamics of BBHs near the scattering threshold? Zoom-whirl orbits are a result of a relativistic gravitational theory. Does the Coulomb repulsion of like-charged BHs suppress them? If zoom-whirl orbits are possible with charged BBHs, are there any new effects in this highly transient and non-linear regime as the charge-to-mass ratio increases? 

Hand-in-hand with this is the question, does charge change the value of \(b_{\rm scat}\)? With an additional channel for energy and angular momentum loss, could the scattering threshold impact parameter be larger for charged BHs? Conversely, could the additional Coulomb force, repulsive for like-charged BHs, lead to a smaller scattering threshold because the binary may have to go deeper in the gravitational potential well to experience stronger relativistic gravity? 
The repulsive force would give the binaries an additional ``push" not to merge.

In a companion paper~\cite{MSmith_paper_II_2024}, which will be referred to as Paper II throughout this work, we broach questions on the extremality of merger remnants from charged BBHs near the scattering regime, and the energy and angular momentum radiated by these mergers. Both this work and Paper II investigate whether the mantra ``matter doesn't matter" \cite{hooft_graviton_1987,amati_superstring_1987,banks_model_1999,east_ultrarelativistic_2013} holds in high-energy interactions of charged BBHs with non-zero impact parameter as it holds in the case of head-on collisions \cite{bozzola_does_2022}. This conjecture states that the internal structure of a BH --- including its charge and/or spin --- has negligible effect at the highest energies. Does this hold near the scattering threshold for charged BHs? Answering these questions requires simulations in full Einstein-Maxwell theory, and here we begin to tackle them.

We conduct simulations of uncharged and like-charged BBHs in full Einstein-Maxwell theory near the scattering threshold. Each BH has an initial boost of $\gamma = 1.520$ ($v/c = 0.753$) for comparison to previous uncharged studies \cite{sperhake_cross_2009}.
We consider initial charge-to-mass ratios \(\lambda \in \{0.0, 0.1, 0.4, 0.6\}\). We also vary the impact parameter $b$. A summary of our results follows.  

We find that zoom-whirl orbits are not suppressed in charged BBHs, as we demonstrate they can arise for binaries with all values of $\lambda$ probed here. We observe no unexpected effects in the dynamics of the zoom-whirl systems. While the qualitative binary dynamics in the zoom-whirl regime remain unchanged as $\lambda$ is varied, the value of the immediate merger threshold impact parameter, $b^*/M_{\rm ADM}$, decreases with $\lambda$. Here $M_{\rm ADM}$ is the Arnowitt-Deser-Misner (ADM) mass. The value of the scattering threshold impact parameter, $b_{scat}/M_{\rm ADM}$, also decreases with $\lambda$. When  normalizing the impact parameters by the sum of the initial irreducible masses $M_{\rm irr}$, we discover universality with charge, i.e., that the values of $b^*/M_{\rm irr}$ and $b_{scat}/M_{\rm irr}$, which determine the location of the zoom-whirl regime in impact parameter space, become independent of $\lambda$. 

These results suggest that the areal radius of the black hole horizon (encoded in the irreducible mass) is a fundamental length scale governing extreme BH encounters in horizon scale scattering experiments.

The fact that charge changes $b_{\rm scat}/M_{\rm ADM}$ and $b^*/M_{\rm ADM}$ demonstrates that charge matters near the scattering threshold for our initial Lorentz factor of $1.52$. Although this Lorentz factor is not in the ultrarelativistic regime, the head-on collisions of~\cite{bozzola_does_2022} showed that charge plays practically no role even at such moderate Lorentz factors. It remains an open question as to whether charge will have an impact at larger initial Lorentz factors. We plan to explore this in future work.

The paper is structured as follows: in Sec. \ref{sec:Methods} we describe our initial data, evolution approach, and diagnostics we adopt. We detail the results of our simulations in Sec. \ref{sec:Results}. We conclude this work in Sec. \ref{sec:Conclusions}. Throughout this paper we adopt geometrized units in which $G=c=(4\pi\epsilon_0)^{-1}=1$, where $G$ is the gravitational constant, $c$ the speed of light, and $\epsilon_0$ the vacuum permittivity.

\section{\label{sec:Methods}Methods}

\subsection{\label{subsec:ID}Initial Data}

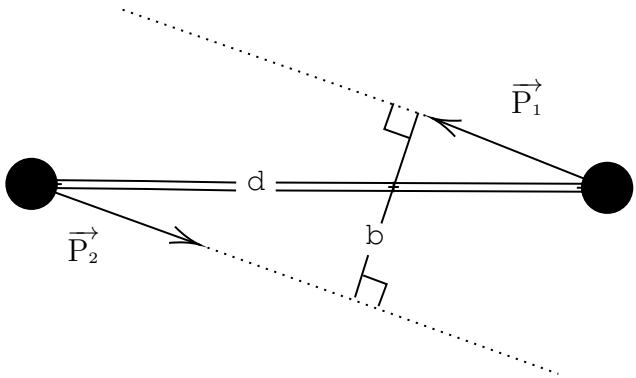
\begin{figure}
\centering
\tikzset{every picture/.style={line width=0.75pt}} 

\begin{tikzpicture}[x=0.999pt,y=0.999pt,yscale=-1,xscale=1]

\draw  [color={rgb, 255:red, 0; green, 0; blue, 0 }  ,draw opacity=1 ][fill={rgb, 255:red, 0; green, 0; blue, 0 }  ,fill opacity=1 ] (219,110) .. controls (219,104.75) and (223.25,100.5) .. (228.5,100.5) .. controls (233.75,100.5) and (238,104.75) .. (238,110) .. controls (238,115.25) and (233.75,119.5) .. (228.5,119.5) .. controls (223.25,119.5) and (219,115.25) .. (219,110) -- cycle ;
\draw  [color={rgb, 255:red, 0; green, 0; blue, 0 }  ,draw opacity=1 ][fill={rgb, 255:red, 0; green, 0; blue, 0 }  ,fill opacity=1 ] (437,111.5) .. controls (437,106.25) and (441.25,102) .. (446.5,102) .. controls (451.75,102) and (456,106.25) .. (456,111.5) .. controls (456,116.75) and (451.75,121) .. (446.5,121) .. controls (441.25,121) and (437,116.75) .. (437,111.5) -- cycle ;
\draw    (305.97,112.5) -- (271.99,111.93) -- (228.49,111.5)(306.03,109.5) -- (272.03,108.93) -- (228.51,108.5) ;
\draw    (321.01,109.5) -- (446.51,110)(320.99,112.5) -- (446.49,113) ;
\draw    (228.5,110) -- (290.12,132.32) ;
\draw [shift={(292,133)}, rotate = 199.91] [color={rgb, 255:red, 0; green, 0; blue, 0 }  ][line width=0.75]    (10.93,-3.29) .. controls (6.95,-1.4) and (3.31,-0.3) .. (0,0) .. controls (3.31,0.3) and (6.95,1.4) .. (10.93,3.29)   ;
\draw    (446.5,111.5) -- (381.86,85.74) ;
\draw [shift={(380,85)}, rotate = 21.73] [color={rgb, 255:red, 0; green, 0; blue, 0 }  ][line width=0.75]    (10.93,-3.29) .. controls (6.95,-1.4) and (3.31,-0.3) .. (0,0) .. controls (3.31,0.3) and (6.95,1.4) .. (10.93,3.29)   ;
\draw  [dash pattern={on 0.84pt off 2.51pt}]  (292,133) -- (428,182) ;
\draw  [dash pattern={on 0.84pt off 2.51pt}]  (263,44) -- (380,85) ;
\draw    (356,137) -- (351,153) ;
\draw    (375,83) -- (361,125) ;
\draw   (354.1,144.68) -- (363,148) -- (359.9,156.32) ;
\draw   (371.89,92.71) -- (362.11,89.24) -- (365.59,79.43) ;

\draw (721,21) node    {$0$};
\draw (313.5,108.75) node   [align=left] {{\fontfamily{pcr}\selectfont \large d}};
\draw (354,124) node [anchor=north west][inner sep=0.75pt]   [align=left] {{\fontfamily{pcr}\selectfont \large b}};
\draw (409,70.4) node [anchor=north west][inner sep=0.75pt]    {$\overrightarrow{\text{\large P}_{1}}$};
\draw (241,126.4) node [anchor=north west][inner sep=0.75pt]    {$\overrightarrow{\text{\large P}_{2}}$};

\draw [draw opacity=0]  (236,110.09) -- (240,110.09)(238,108.09) -- (238,112.09) ;
\draw [draw opacity=0]  (226.5,110) -- (230.5,110)(228.5,108) -- (228.5,112) ;
\draw [draw opacity=0]  (435,111.46) -- (439,111.46)(437,109.46) -- (437,113.46) ;
\draw [draw opacity=0]  (444.5,111.5) -- (448.5,111.5)(446.5,109.5) -- (446.5,113.5) ;
\draw [draw opacity=0]  (363.61,111.18) -- (367.61,111.18)(365.61,109.18) -- (365.61,113.18) ;
\end{tikzpicture}
\caption{\label{fig:setup_diag}A diagram depicting the set-up for our simulations, with initial coordinate separation $d$, impact parameter $b$, and BH initial momenta $P_{1,2} = (\mp P_x, \pm P_y, 0)$. }
\end{figure}

The setup of our simulations follows closely that of \cite{sperhake_cross_2009}, for comparison to their uncharged study. Two equal-mass, equal-charge, nonspinning BHs are separated by an initial coordinate separation \(d\) along the x-axis in the x-y plane. The BHs are boosted by the same Lorentz factor $\gamma$ and have initial momenta \(P_{1,2} = (\mp P_x, \pm P_y, 0)\). An impact parameter $b$ describes the perpendicular coordinate distance between the two initial trajectories. A geometric figure of the set-up is shown in Fig.~\ref{fig:setup_diag}. 

\begin{table*}
\caption{\label{table:zw_ID} Initial data parameters for the binaries in our study, which have initial linear momentum $\left|P \right| =  0.57236$ corresponding to a Lorentz factor of approximately 1.520. $\overline{\lambda}$ is the initial charge-to-mass ratio of each BHs, $\overline{\lambda} \equiv Q/ M$, computed via the isolated horizon formalism,  and $\lambda$ is the target initial charge-to-mass ratio which we use to label the initial data in this work. $M_{\rm ADM}$ is the spacetime ADM mass; $M$ and $M_{\rm irr}$ are the sums of the initial quasilocal masses  and irreducible masses of the 
BHs, respectively. $b/M_p$ and $d/M_p$ are the values input into \texttt{TwoChargedPunctures} with $M_p=1$ in code units. We also list the value of the impact parameter $b$ normalized to $M_{\rm ADM}$ and $M_{\rm irr}$. In the second-to-last column, $a\equiv J_{\rm ADM}/M_{\rm ADM}$, where $J_{\rm ADM}$ is the spacetime ADM angular momentum. }
\begin{ruledtabular}
\begin{tabular}{cccccccccc}
$\lambda$ & $\overline{\lambda}$ & $M_{\rm ADM}$& $M$ & $M_{\rm irr}$ & $b/M_p$ & $b/M_{\rm ADM}$ & $b/M_{\rm irr}$ & $a/M_{\rm irr}$ & $d/M_{\rm ADM}$\\
\colrule
0.0 & 0.000 & 1.543 & 1.001 & 1.001 & 5.153 & 3.340 & 5.148 & 1.907 & 171.536 \\
0.1 & 0.100 & 1.543 & 1.001 & 0.999 & 4.998 & 3.239 & 5.005 & 1.854 & 171.480 \\
0.1 & 0.100 & 1.543 & 1.001 & 0.999 & 5.075 & 3.289 & 5.083 & 1.882 & 171.480 \\
0.1 & 0.100 & 1.543 & 1.001 & 0.999 & 5.114 & 3.314 & 5.122 & 1.897 & 171.480 \\
0.1 & 0.100 & 1.543 & 1.001 & 0.999 & 5.134 & 3.327 & 5.141 & 1.904 & 171.480 \\
0.1 & 0.100 & 1.543 & 1.001 & 0.999 & 5.153 & 3.339 & 5.161 & 1.911 & 171.480 \\
0.4 & 0.399 & 1.551 & 1.002 & 0.961 & 3.607 & 2.325 & 3.755 & 1.384 & 170.613 \\
0.4 & 0.399 & 1.551 & 1.002 & 0.961 & 4.766 & 3.073 & 4.962 & 1.828 & 170.613 \\
0.4 & 0.399 & 1.551 & 1.002 & 0.961 & 4.863 & 3.135 & 5.063 & 1.865 & 170.613 \\
0.4 & 0.399 & 1.551 & 1.002 & 0.961 & 4.911 & 3.166 & 5.113 & 1.884 & 170.613 \\
0.4 & 0.399 & 1.551 & 1.002 & 0.961 & 4.936 & 3.182 & 5.138 & 1.893 & 170.613 \\
0.4 & 0.399 & 1.551 & 1.002 & 0.961 & 4.948 & 3.190 & 5.151 & 1.898 & 170.613 \\
0.4 & 0.399 & 1.551 & 1.002 & 0.961 & 4.960 & 3.198 & 5.164 & 1.902 & 170.613 \\
0.4 & 0.399 & 1.551 & 1.002 & 0.961 & 5.153 & 3.322 & 5.365 & 1.977 & 170.613 \\
0.1 & 0.100 & 1.526 & 1.001 & 0.999 & 4.980 & 3.263 & 4.987 & 1.868 & 62.151 \\
0.4 & 0.399 & 1.534 & 1.002 & 0.960 & 4.780 & 3.116 & 4.978 & 1.855 & 61.834 \\
0.6 & 0.598 & 1.545 & 1.003 & 0.904 & 4.507 & 2.916 & 4.987 & 1.844 & 61.377 \\
0.6 & 0.598 & 1.545 & 1.003 & 0.904 & 4.540 & 2.938 & 5.024 & 1.858 & 61.377
\end{tabular}
\end{ruledtabular}
\end{table*}

To compare with previous uncharged studies and across initial parameters, we keep the isolated-horizon quasilocal mass~\cite{dreyer_introduction_2003,bozzola_initial_2019} fixed to very high precision for a given $\lambda$, but because of the way we setup our initial data, this is achieved to within $O(0.1\%)$ for different values of $\lambda$.

The initial data for our setup are generated by \texttt{TwoChargedPunctures}~\cite{bozzola_initial_2019}, which provides the metric and electromagnetic vector potential for two punctures given bare properties of the BHs. In particular, we specify the initial linear momenta $P_p^{1,2}$ and charges $Q_p^{1,2}$ of the BHs, and set target quasilocal gravitational masses $M_p^{1,2}$ (see \cite{ansorg_single-domain_2004}). \texttt{TwoChargedPunctures} allows for generic BH spin angular momenta, but our BHs are all initially nonspinning. We choose $M_p^{1,2} = 0.5$ and $Q^{1,2}_p \in \{0.0, 0.05, 0.2, 0.3\}$ to target charge-to-mass ratios $\lambda \in \{0.0, 0.1, 0.4, 0.6\}$. We set the same target $\lambda$ for the two BHs in a binary, and define $M_p \equiv M_p^1 + M_p^2 = 1.0$. The quasilocal gravitational masses $M^1 = M^2$ of the BHs are close to but do not perfectly coincide with $M_p^{1,2} = 0.5$. For all values of $b$, $M^{1,2}$ of $\lambda = 0.0$ binaries differ from the target value of $0.5$ by $0.1\%$, and $M^{1,2}$ for $\lambda = 0.4, 0.6$ binaries differ from the target by $0.2 \%$ and $0.3 \%$, respectively. For a given $\lambda$, $M^{1,2}$ remains fixed to within 2 parts in $10^5$ as $b$ is varied.  

While the quasilocal gravitational BH masses have slight variation, the magnitude of the linear momentum, 
\begin{eqnarray} \label{eq:Pmag}
\left| P \right| \equiv \sqrt{(P_x)^2 + (P_y)^2} = M^{1,2} \sqrt{\gamma^2 - 1},
\end{eqnarray}
is held constant for all simulations at $|P| = 0.57236$, which corresponds to a target $\gamma = 1.520$ for $M_p^{1,2} = 0.5$. The slight variation in the quasilocal gravitational mass is compensated for by small variations in the initial Lorentz factor, so our initial data correspond to quasilocal BH masses and initial Lorentz factors fixed to within $O(0.1\%)$. The values for $P_p^{1,2} = (\mp P_x, \pm P_y, 0)$ are calculated as follows
\begin{eqnarray} 
P_y = \left|P\right| \frac{b}{d} \, ; \,\,\,\,\,\,\, P_x = \left|P\right| \sqrt{1 - \left(\frac{b}{d}\right)^2}\, 
\label{eq:PxPy}.
\end{eqnarray}
The punctures are located at $X_p^{1,2} = (\pm d/2, 0, 0)$. 

For \(\lambda \in \{0.1, \, 0.4\}\) in our suite of simulations, we investigate the effects of impact parameter while keeping $\left| P \right|$ constant. For consistency, the \(\lambda \in \{0.0, \,  0.1, \, 0.4\}\)  binaries have an initial separation of \(d/M_p = 264.63\). To demonstrate that zoom whirl behavior can take place for higher $\lambda$ and different initial separations, we also present a case with \(d/M_p = 94.85\) and $\lambda=0.6$, and include in our dataset cases with $\lambda=0.1, 0.4$ at \(d/M_p = 94.85\) which also exhibit zoom-whirl behavior. Unless otherwise noted, the conclusions and analyses in this work are performed on the binaries with initial coordinate separation $d/M_p = 264.63$. A detailed description of initial data parameters for our simulations is presented in Table \ref{table:zw_ID}. Outside of Table \ref{table:zw_ID}, we label the BHs by their target initial charge-to-mass ratios $\lambda$.

\begin{figure*}
\includegraphics[width=0.95\textwidth]{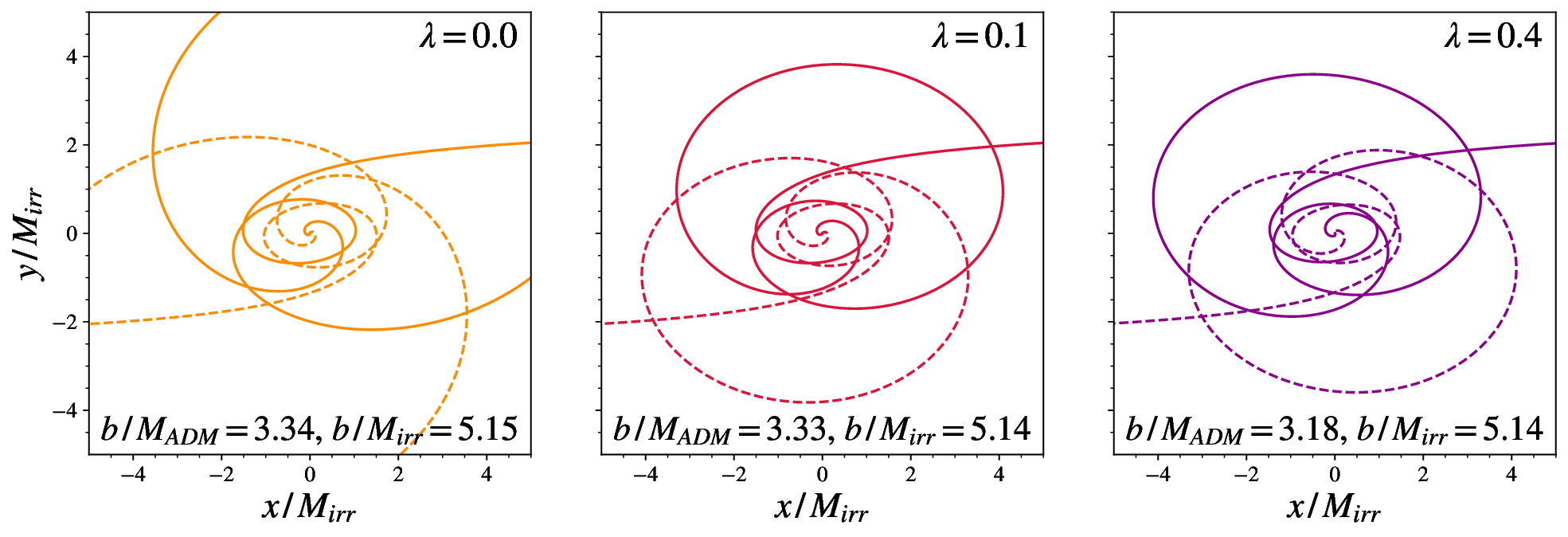}
\caption{\label{fig:zw_q_traj} 
Representative puncture trajectories from simulations with zoom-whirl orbits (left to right: \(\lambda = 0.0, \, 0.1, \, 0.4\)). We did not include an investigation of \(\lambda = 0.0\) with impact parameter, but a single simulation with an impact parameter known to induce zoom-whirl orbits from \cite{sperhake_cross_2009} was run for comparison.}
\end{figure*}

\subsection{\label{subsec:evol}Spacetime Evolution}

Our simulations are conducted with the \texttt{Einstein Toolkit} \cite{brandt_einstein_2021} and analyzed with \texttt{kuibit} \cite{bozzola_kuibit_2021}. The Einstein-Maxwell equations are solved in the 3+1 decomposition (e.g.~\cite{arnowitt_dynamics_2008,baumgarte_ch_2010}), using the \texttt{LeanBSSNMoL} thorn~\cite{sperhake_binary_2007} for the spacetime and the massless version of \texttt{ProcaEvolve} \cite{zilhao_nonlinear_2015} for the electromagnetic field. These codes are part of the \texttt{Canuda} suite \cite{witek_canuda_2021} which have been tested extensively for convergence and other tests, and {\tt LeanBSSNMoL} was also used in previous high-energy BH scattering simulations, e.g., in~\cite{sperhake_cross_2009, sperhake_universality_2013}. The Einstein equations are solved in the Baumgarte-Shapiro-Shibata-Nakamura formalism \cite{shibata_evolution_1995, baumgarte_numerical_1998}. We employ the 1+log slicing condition and Gamma-driver gauge condition with the shift damping parameter $\eta$ (see Eq.~17 of \cite{hinder_error-analysis_2013}) set to $1.0/M_p$ for the spacetime evolution, and the Lorenz condition for the electromagnetic gauge field. We adopt \nth{6}-order finite differencing for spatial derivatives, and the time integration is carried out using the \nth{4}-order Runge-Kutta method.

We use Cartesian grids with adaptive mesh refinement provided by \texttt{Carpet} \cite{schnetter_evolutions_2004}. Each simulation has three sets of nested refinement levels, two of which remain centered on the punctures and one on the origin. We have \(10\) or $11$ refinement levels, depending on the initial charge-to-mass ratio of the binary and the initial coordinate separation of the BHs. 
Our resolution varies with the charge-to-mass ratio of the binary to adjust for the change in horizon size. The resolution is chosen such that the smallest coordinate radius of the BH apparent horizons is resolved by \(\sim 33\) grid-points after the initial data relaxes to the evolution gauge. This corresponds to finest resolutions of $ \{ $$M_p/89, \,$$ M_p/91, \, M_p/98 \,, M_p/114$ $\}$ for \(\lambda = \{ 0.0, \, 0.1, \, 0.4,\, 0.6\}\), respectively.  
We place the outer grid boundary at \(\sim 1000 M_p\), with the exception of the $\lambda = 0.1$, $b/M_p = 5.153$ binary, for which the boundary is placed at $535 M_p$. For the binaries with smaller initial coordinate separation, our outer grid boundary is placed at least $500 M_p$ from the origin for $\lambda = 0.1$ and at least $850 M_p$ from the origin for $\lambda = 0.4, 0.6$. Larger $\lambda$ binaries typically evolve more slowly due to the electrostatic repulsion as seen also in~\cite{bozzola_numerical-relativity_2021,bozzola_can_2023}. For this reason we choose the outer boundary further out in larger $\lambda$ cases to ensure the boundary does not come into causal contact with the binary. In all cases the outer boundary is causally disconnected for the length of our evolutions. We use the Kreiss-Oliger dissipation prescription as introduced in \cite{bozzola_numerical-relativity_2021}, but proportional to the local Courant factor.

The computational cost for a simulation exhibiting zoom-whirl orbits, can exceed $75,000$ CPU hours. This translates to more than $330$ hours of wall-clock time per simulation, excluding queue waiting time.

\subsection{\label{subsec:diag}Diagnostics}

We use \texttt{AHFinderDirect}~\cite{thornburg_fast_2003} to locate apparent horizons and \texttt{QuasiLocalMeasuresEM}~\cite{bozzola_initial_2019} -- our version of the original \texttt{QuasiLocalMeasures}~\cite{dreyer_introduction_2003} that accounts for charge -- to calculate BH quasilocal diagnostics. \texttt{QuasiLocalMeasuresEM} computes the charge, angular momentum, gravitational mass, and irreducible mass of a BH via the isolated horizon formalism. We extract gravitational and electromagnetic radiation with the Newman-Penrose formalism via \texttt{NPScalarsProca} \cite{newman_approach_1962, witek_black_2010, zilhao_nonlinear_2015} (see \cite{bishop_extraction_2016} for a review on extraction of gravitational waves in numerical relativity). Our Weyl Newman-Penrose scalar $\Psi_4$ 
is reported at $r/M_p = 80$.

The ADM mass, and ADM angular momentum, $J_{\rm ADM}$, of the spacetime are calculated by \texttt{TwoChargedPunctures} at the initial data level. The initial irreducible masses, gravitational masses, and charges of the BHs are calculated by \texttt{QuasiLocalMeasuresEM} with Eqs.~A21, A23, and A20, respectively, of \cite{bozzola_initial_2019}. These BH properties vary slightly as the initial data relax, so we extract them at $t/M_{\rm ADM} = 19.5$, after initial data relaxation, which are the values reported in Table~\ref{table:zw_ID}. 

\section{\label{sec:Results}Results}

\subsection{Observation of Zoom-Whirl Orbits}

A key result from our study is that charged BBHs can exhibit zoom-whirl orbits independently of the charge-to-mass ratios we explored. Figure~\ref{fig:zw_q_traj} displays representative zoom-whirl puncture trajectories from binaries with initial charge-to-mass ratios \(\lambda = \{0.0, 0.1, 0.4\}\). The zoom-whirl trajectory for $\lambda = 0.0$ is included for comparison. The characteristic ``whirl" phase where the punctures have a close encounter is followed by the punctures ``zoom"-ing out to more than twice their whirl-phase coordinate separation. We see no qualitative differences between the dynamics of the zoom-whirl orbits exhibited by the charged and uncharged binaries. However, there are quantitative differences in the threshold impact parameters that give rise to zoom-whirl orbits that we discuss below.

\begin{figure}
\includegraphics[width=0.33\textwidth]{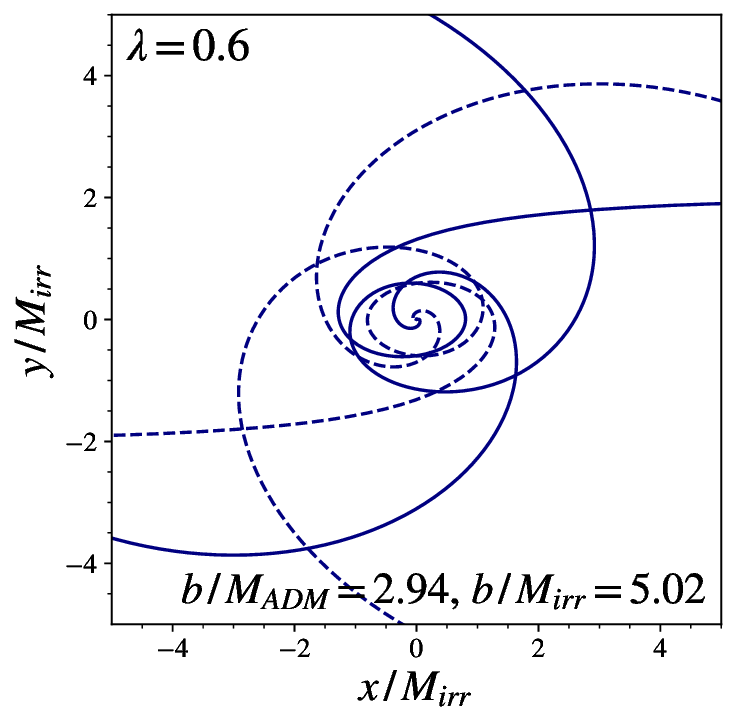}
\caption{\label{fig:zw_q6} Puncture trajectories for a $\lambda = 0.6$ binary that exhibits zoom-whirl orbits. The linestyles differentiate the two BHs.}
\end{figure}

Among the cases listed in Table \ref{table:zw_ID}, we include binaries for $\lambda = 0.1, 0.4, 0.6$ with smaller initial coordinate separations. These binaries are included because their impact parameters fall in the zoom-whirl regime. These binaries have multiple encounters prior to merger as identified with the method described in Sec. \ref{subsubsec:threshold_calc}, but most do not separate to twice the whirl-phase coordinate separation between the first BH encounter and merger. The exception is the $\lambda = 0.6$ binary whose puncture trajectories (and impact parameter) are depicted in Fig.~\ref{fig:zw_q6}. The BHs in this binary merge $t \sim 230 M_{\rm ADM}$ after their first encounter.

These results demonstrate that zoom-whirl behavior can be exhibited for at least $\lambda=0.6$, and that the electrostatic repulsion does not suppress zoom-whirl orbits. In a future work, we plan to probe even larger $\lambda$ to test the limits of zoom-whirl behavior.

%

\subsection{Impact Parameter Thresholds and Universality With the Irreducible Mass}

For each charge-to-mass ratio, a threshold impact parameter $b_{\rm scat}$ separates the impact parameters for which binaries ultimately merge ($b<b_{\rm scat}$) and those for which binaries scatter to infinity ($b>b_{\rm scat}$). Additionally, an immediate merger threshold $b^*<b_{\rm scat}$ separates the impact parameters that result in binaries merging within the first encounter ($b < b^*$) from impact parameters that result in multiple BH encounters ($b^* < b < b_{\rm scat}$) before merging \cite{pretorius_black_2007, sperhake_cross_2009}. The zoom-whirl orbits pictured in Figs.~\ref{fig:zw_q_traj},~\ref{fig:zw_q6} are examples of binaries with multiple encounters prior to merger. By densely sampling in impact parameter space and using the methods we describe in the next section, we identify
$b_{\rm scat}$ and $b^*$ for $\lambda = 0.1$, and for $\lambda =0.4$. 

\begin{table}
\caption{\label{tab:zw_bvals}
Estimates of \(b^*\) and \(b_{\rm scat}\) for the two $\lambda$. When normalized by \(M_{\rm irr}\), the respective values are universal to within the determination error across $\lambda$. We include $\overline{b} \equiv J_{\rm ADM}/\left|P\right|$ as another estimate of the impact parameter.}
\begin{ruledtabular}
\begin{tabular}{ccc}
 $\lambda$ & 0.1 &0.4  \\
 \hline
 $b^*/M_{\rm ADM}$ & $3.30 \pm 0.01$ & $3.15 \pm 0.02$  \\
  $b^*/M_{\rm irr} $ & $5.10 \pm 0.02$ &$5.09 \pm 0.03$ \\
  $\overline{b}^*/M_{\rm ADM}$ & $3.30 \pm 0.01$ & $3.15 \pm 0.02$  \\
  $\overline{b}^*/M_{\rm irr} $ & $5.09 \pm 0.02$ &$5.08 \pm 0.03$ \\
$b_{\rm scat}/M_{\rm ADM}$  & $3.333 \pm 0.006$ & $3.194 \pm 0.004$ \\
 $b_{\rm scat}/M_{\rm irr}$  &$5.15 \pm 0.01$ & $5.157 \pm 0.006$ \\
 $\overline{b}_{\rm scat}/M_{\rm ADM}$  & $3.328 \pm 0.006$ & $3.189 \pm 0.004$ \\
 $\overline{b}_{\rm scat}/M_{\rm irr}$  &$5.14 \pm 0.01$ & $5.149 \pm 0.006$ \\
\end{tabular}
\end{ruledtabular}
\end{table}

Table~\ref{tab:zw_bvals} lists our results for the values of these thresholds. We find that the values of both $b_{\rm scat}/M_{\rm ADM}$ and $b^*/M_{\rm ADM}$ decrease with larger initial charge-to-mass ratio; the repulsive Coulomb force of the like-charged BHs modifies the location of the zoom-whirl regime. We notice that both $b_{\rm scat}/M_{\rm ADM}$ and $b^*/M_{\rm ADM}$ for $\lambda=0.1$ are 4-5\% larger than for $\lambda=0.4$ despite that $\lambda=0.4$ is not even close to extremal. This is a second key finding in our study: For the Lorentz factors probed here, head-on collisions of charged black holes with $\lambda \leq 0.8$ behave like the uncharged 
counterpart~\cite{bozzola_does_2022}. Unlike head-on collisions, we observe that charge has measurable effects near the scattering threshold regime for our initial Lorentz factor of $\sim 1.5$. If charge does not matter for high-energy scattering events, the energy scales for that to be true would have to be higher than what is probed in this work. In Paper II we will provide additional metrics which demonstrate that charge has a more dramatic impact in this regime for Lorentz factors of $\simeq 1.5$.

A third key finding in our work is that while the threshold impact parameters are charge-dependent when normalized by $M_{\rm ADM}$, we find that $b_{\rm scat}/M_{\rm irr}$ and $b^*/M_{\rm irr}$ become universal, i.e., independent of $\lambda$, for the values probed here. Here we remind the reader that $M_{\rm irr}$ is the sum of the initial irreducible masses. In particular, for $\lambda = 0.1, 0.4$ we obtain \( 5.08 \leq b^*/M_{\rm irr} \leq 5.12\) and \(5.15 \leq b_{\rm scat}/M_{\rm irr} \leq 5.16 \), respectively (see Table~\ref{tab:zw_bvals}). These results are in remarkable agreement with~\cite{sperhake_cross_2009}, who found $b^*/M_{\rm irr} = 5.09 \pm 0.02$ and $ b_{\rm scat}/M_{\rm irr} \leq 5.2$ when we convert their published $b/M_{\rm ADM}$ values to $b/M_{\rm irr}$ using their $\gamma = 1.520$.  In Table~\ref{tab:zw_bvals} we also include another estimate for the impact parameter: $\overline{b} \equiv J_{\rm ADM}/\left|P\right|$. Since the BHs are initially non-spinning, $J_{\rm ADM}$ is the orbital angular momentum of the spacetime that motivates the above definition. The results in the table, show that thresholds for scattering and immediate merger measured with $\overline{b}$, also become universal, when $\overline{b}$ is normalized to $M_{\rm irr}$.

In Paper II, we show that universality with the irreducible mass extends to quantities related to merger remnants as well, and that the values are universal for $\lambda$ up to $0.6$. 

We note that the impact parameters $b/M_{\rm irr}$ for the binaries with smaller initial coordinate separations do not fall in the zoom-whirl range predicted by the $\lambda=0.1$ and $0.4$ cases at the larger separation we systematically studied. This is a result of the different initial separation at which the binaries start, and possibly also the definition of impact parameter at finite separation. We will expand more on this topic in Paper II.
 
In our effort to discover an impact parameter normalization and scaling with $\lambda$ that allows us to predict the threshold impact parameters independently of $\lambda$, we tested a range of impact parameter normalizations on $b_{\rm scat}$ for $\lambda = 0.1$ and $\lambda = 0.4$.\footnote{We use $b_{\rm scat}$ because we identified it more accurately than $b^*$.} The scalings we explored include $\left( 1 - \lambda^2 \right)$, $\left(\gamma^2 - \lambda^2 \right)$, and $ \sqrt{1 - \lambda^2}$, with $b$ normalized by $M_{\rm ADM}$, $M_{\rm irr}$, and $M_1$ ($M_1$ is the initial gravitational mass of one of the binary's BHs). In these calculations we use the precise initial charge-to-mass ratio $\overline{\lambda}$, calculated after initial data relaxation. The $(1 - \lambda^2)$ and $\sqrt{1 - \lambda^2}$ factors are motivated by studies of head-on collisions of charged BBHs~\cite{zilhao_collisions_2012, bozzola_does_2022}. Reference~\cite{zilhao_collisions_2012} found that when the BHs started from rest, collision times scaled like $t \sim 1/ \sqrt{1 - \lambda^2}$, and the waveforms scaled with $(1 - \lambda^2$). Reference~\cite{bozzola_does_2022} investigated moderate boosts ($\gamma < 1.5$) and found that these scalings did not hold. The scaling of $(\gamma^2 - \lambda^2)$ is motivated by a Newtonian-like expression for the sum of the electrostatic and gravitational forces:
\begin{eqnarray}
F_{tot} = - \frac{M^2}{r^2}\left[\gamma^2 - \lambda^2 \right]
\label{eq:Newton_approx},
\end{eqnarray}
where $M$ is the gravitational mass of the BHs.

Among these scalings and normalizations, only $b_{\rm scat}/M_{\rm irr}$ yields a value almost independent of $\lambda$ for both $\lambda = 0.1$ and $\lambda = 0.4$, with $5.15 \leq b_{\rm scat}/M_{\rm irr} \leq 5.16$ to within determination error. Note that the  irreducible mass of an unboosted Reissner-Nordstr\"om BH, which would describe our initially nonspinning BHs (if they were , involves both the gravitational mass $M$ and $\lambda$:
\begin{eqnarray}
M_{\rm irr} = \frac{M}{2} \left[ 1 + \sqrt{1 - \lambda^2} \right]
\label{eq:mirr_formula}.
\end{eqnarray}

Our results suggest that the areal radius $r_A=\sqrt{A/4\pi}=2M_{\rm irr}$ of BH horizons (encoded in the irreducible mass $A=16\pi M_{\rm irr}^2$) is a fundamental quantity governing extreme BH encounters in horizon scale scattering experiments. In a follow-up work~\cite{MSmith_paper_III}, we will test the limit of this universality with higher values of $\lambda$ and in the case of BHs with non-zero spin, including larger values of the initial Lorentz factor.

\subsection{\label{subsubsec:threshold_calc}Identification of Impact Parameter Thresholds}

To determine the scattering threshold for each $\lambda$, we first group the binaries by whether they merge or scatter as we vary the impact parameter. The scattering binaries are confirmed via analysis of the binary binding energy after the BHs have separated by at least $r = 25 M_{\rm ADM}$ and at least $t = 150 M_{\rm ADM}$ after the first close encounter. For a given $\lambda$, we identify $b_{\rm scat}$ with the average of the largest impact parameter that leads to a merger and the smallest impact parameter that leads to scattering. The error reported with our results in Table \ref{tab:zw_bvals} is the distance $\Delta b$ from the mean to the two values of $b$ used to determine $b_{\rm scat}$. This process is repeated for each $\lambda$ in our set.

Although puncture trajectories have the most dramatic indication of zoom-whirl behavior, they are gauge-dependent. Therefore, we define whether a binary is above or below the immediate merger threshold by examination of the GW signal. 
There are three periods of gravitational radiation in a generic BBH merger: inspiral, merger, and ringdown. Following merger, the quasinormal modes of the ringdown phase are described by \(\Psi_4  \propto e^{-i \omega t}\), where $\Psi_4$ is the Newman-Penrose scalar encoding outgoing gravitational radiation, and $\omega$ the complex ringdown frequency~\cite{teukolsky_rotating_1972}. The log of the magnitude of \(\Psi_4\) in this phase is a linear curve with respect to time.

\begin{figure}
\includegraphics[width=0.47\textwidth]{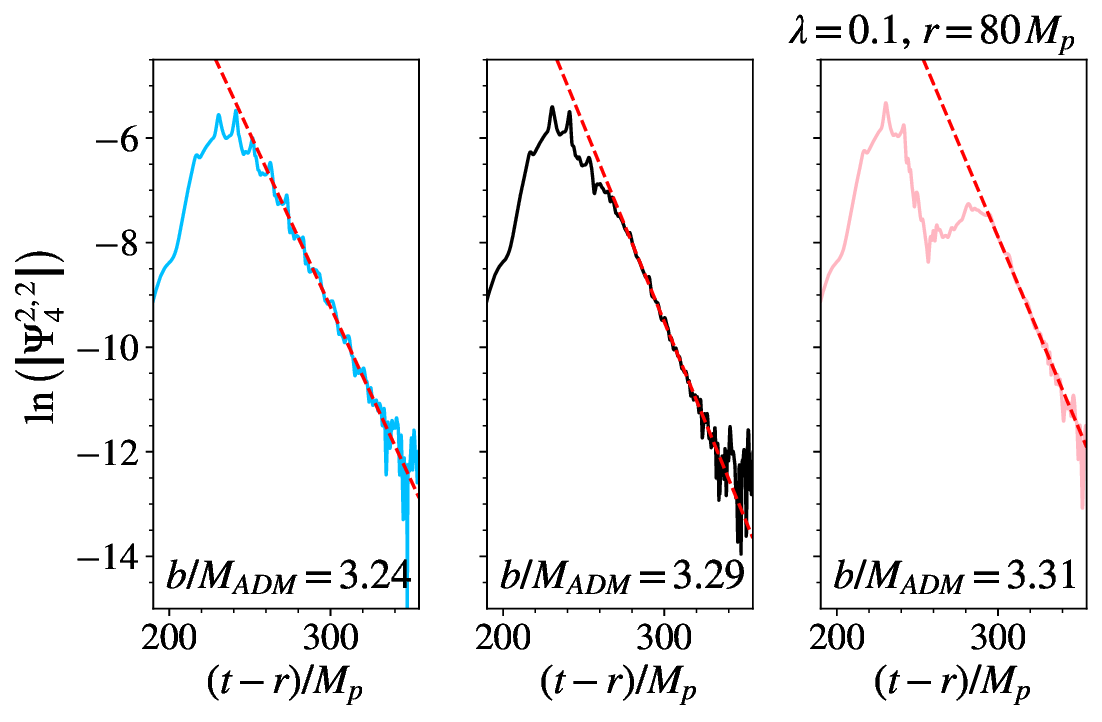}
\caption{\label{fig:GW_bstar} The natural log of the magnitude of \(\Psi^{(2,2)}_4\) for three $\lambda = 0.1$ binaries near the threshold of immediate merger. The red dashed lines are least squares fits to the linear curve of the ringdown phase.} 
\end{figure}

Figure \ref{fig:GW_bstar} plots the natural log of the $\ell=2,m=2$ mode of the amplitude of $\Psi_4$ (\(\Psi^{2,2}_4 \)) for three $\lambda = 0.1$ binaries near the immediate merger threshold.
If the BHs have only one encounter prior to merger, the time evolution of \(\ln \left|\Psi_4 \right| \) exhibits a single peak and a smooth transition to a linear curve (e.g. Fig. \ref{fig:GW_bstar} [left], $b/M_{\rm ADM} = 3.24$). If the BHs have more than one encounter prior to merger, \(\ln \left|\Psi_4 \right| \) will exhibit at least two peaks prior to ringdown (e.g. Fig. \ref{fig:GW_bstar} [right], $b/M_{\rm ADM} = 3.31$). The $b/M_{\rm ADM} = 3.29$ binary in Fig. \ref{fig:GW_bstar} [center] does not have the clear repeating features of $b/M_{\rm ADM} = 3.24$, but also no clearly defined secondary peak to declare it past the immediate merger threshold. It is possible this binary demonstrates the blurring of the immediate merger threshold referred to in \cite{sperhake_cross_2009}. Regardless, these GW signals indicate that the immediate merger threshold lies between $b/M_{\rm ADM} = 3.29$ and $b/M_{\rm ADM} = 3.31$ for $\lambda = 0.1$. We thus report $b^*/M_{\rm ADM} = 3.30 \pm 0.01$.

\section{\label{sec:Conclusions}Conclusions}

High-energy black hole collisions are testing grounds for fundamental physics. In particular, these interactions allow us to investigate general relativity coupled to a U(1) gauge theory. In this work, we study high-energy collisions near the scattering threshold of equal-mass black holes endowed with the same charge and fixed initial linear  momenta, which correspond to an almost fixed initial Lorentz factor of 1.52. We demonstrate that the relativistic phenomenon of zoom-whirl orbits can be exhibited by BBHs with initial charge-to-mass ratios of at least $\lambda = 0.6$. 

We find that zoom-whirl behavior in binaries with fixed initial $\gamma=1.520$ occur at lower $b/M_{\rm ADM}$ as $\lambda$ increases. In particular, the values of the scattering and immediate merger impact parameter thresholds decrease with increasing $\lambda$. We conclude that as $b/M_{\rm ADM}$ approaches $b^*/M_{\rm ADM}$, the threshold of immediate merger, and $b_{\rm scat}/M_{\rm ADM}$, the scattering treshold, charge leaves observable imprints in key properties
at energy scales where charge has negligible influence for the head-on case. In a follow-up paper~\cite{MSmith_paper_II_2024} we will provide additional metrics which demonstrate that charge has a more dramatic impact near the scattering threshold regime. 

We investigate impact parameter normalizations for which the scattering thresholds might agree across $\lambda$, and find that only $b_{scat}/M_{\rm irr}$ produces universal behavior, i.e., insensitive to $\lambda$, where $M_{\rm irr}$ is the sum of the initial irreducible masses. We find that $b^*/M_{\rm irr}$ is also independent of $\lambda$. These results indicate the importance of $M_{\rm irr}$, which is proportional to the areal radius of a BH, in setting the scale of the problem. In other words, for horizon scale interactions near the scattering threshold, it should be the relative size of the areal radius to the impact parameter that determines the outcome. 

Our results need to be tested across larger values of $\lambda$ and BHs with spin, as well as unequal mass binaries. If, as suggested in~\cite{pretorius_black_2007}, the majority of the kinetic energy can be radiated away with sufficient fine tuning, then one would expect that charge would eventually matter at any energy scale. Whether this can take place appears to still be a matter of debate~\cite{PhysRevD.107.064057,sperhake_universality_2013}. These will be the topics of an upcoming paper of ours~\cite{MSmith_paper_III}. We defer a convergence study of similar and more challenging simulations than those performed here in our follow-up paper~\cite{MSmith_paper_II_2024}.

\begin{acknowledgments}
We are grateful to the developers and maintainers of the open-source codes that we used. \texttt{Kuibit}~\cite{bozzola_kuibit_2021}, used in our analysis, uses \texttt{NumPy} \cite{harris_array_2020}, \texttt{scipy} \cite{virtanen_scipy_2020}, and \texttt{h5py} \cite{[][{, http://www.h5py.org/.}]noauthor_hdf5_nodate}. \texttt{Matplotlib} \cite{hunter_matplotlib_2007} was used to generate our figures. We thank Vikram Manikantan for his feedback on figures displayed in this work, and Maria Mutz for useful discussions and feedback on the manuscript. This work was in part supported by NSF Grant PHY-2145421 and 
NASA grant 80NSSC24K0771NASA. This work was supported by Advanced Cyberinfrastructure Coordination Ecosystem: Services \& Support (ACCESS) 
allocation TG-PHY190020 and Frontera allocation PHY23009. ACCESS is funded by NSF awards No.~2138259, No.~2138286, No.~2138307, No.~2137603 and No.~2138296 under the Office of Advanced Cyberinfrastructure. The simulations were performed on \texttt{Stampede2} and \texttt{Frontera}, funded by NSF awards No.~1540931 and No.~1818253, respectively, at the Texas Advanced Computing Center (TACC). 
\end{acknowledgments}

\bibliography{references}

\end{document}